\documentclass[12pt,preprint]{aastex}

\shorttitle{Radial Distributions in NGC~6441}
\shortauthors{D.~A. Krogsrud}

\providecommand{\gtsimeq}{\raisebox{-0.6ex}{$\,\stackrel{\raisebox{-.2ex}{$\textstyle >$}}{\sim}\,$}}

\providecommand{\e}[1]{\ensuremath{\times 10^{#1}}}

\begin{document}

\title{The Curious Radial Distributions of Horizontal Branch Stars in NGC~6441}

\author{David A. Krogsrud, Eric L. Sandquist, Tadafumi Kato}
\affil{San Diego State University}
\email{dkrogsru@gemini.edu,erics@sciences.sdsu.edu}

\begin{abstract}
NGC 6441 is one of the most massive and most metal-rich globular clusters in
the galaxy, and is noted for an unusual extended horizontal branch which
reaches past the instability strip.  We find evidence that there are two
different populations of stars {\it within} the heavily-populated red
clump. Once a differential reddening correction is applied, a large but
compact group of stars is found at the faint red end of the clump in the color
magnitude diagram. Brighter, bluer stars in the clump are found to be more
centrally concentrated within the cluster at very high level of significance.
Curiously, the blue horizontal branch stars show a more complex distribution,
and are not more centrally concentrated than the brighter red clump stars.
The spatial distributions of clump stars is in agreement with the idea that
the the brighter bluer part of the clump is a helium-enriched second
generation.  The blue horizontal branch stars may be showing evidence that
they are being dynamically evaporated.
\end{abstract}

\keywords{globular clusters: individual: NGC 6441 --- stars: abundances ---
  stars: horizontal branch --- stars: kinematics and dynamics}

\section{Introduction}

Globular clusters (GCs) have long been seen as examples of simple stellar
populations, though the current puzzle of multiple populations began with the
discovery of ``chemical anomalies'' within GCs over thirty years ago (see
\citealt{kraft} for a review of early studies).  Many studies have shown
that while generally homogeneous in Fe-peak elements, GC stars can exhibit
internal spreads in light element abundances (a more recent review can be
found in \citealt{gratton04}).  These anomalies have been observed at both the
turnoff and along the subgiant branch \citep{gratton01}, and so the stars must
have formed with at least some of the chemical anomalies.

High-precision photometric data have provided the means to identify broader
evidence for multiple populations. \citet{bedin04} used \textit{Hubble Space
  Telescope} (\textit{HST}) data to show that the main sequence (MS) in
$\omega$ Centauri is split into two components.  Surprisingly, the bluer MS
has twice the metallicity of the red component \citep{piotto05}, which can
only be explained by helium enrichment of the more metal-rich component.
\citet{piotto07} found that the MS of NGC 2808 is split into three components,
and concluded this is probably also evidence of stellar populations with
differing helium abundance. Splits on the subgiant branch (SGB) of multiple
clusters have also been reported, including $\omega$ Cen, NGC 1851, NGC 2808,
and NGC 6388 \citep{piotto09}. However, the oldest photometric evidence
involves the unusual distribution of stars along the horizontal branch
(HB). In this area, the clusters NGC 6388 and NGC 6441 are particularly
noteworthy.  Both clusters have high metallicity ($\langle$[Fe/H]$\rangle =
-0.45$ and $-0.44$, respectively; \citealt{carretta09b}), yet have extended
horizontal branches that continue past the instability strip to $T_{eff} >
30,000$ K.  The RR Lyrae variables in both clusters have unusually long
periods \citep{pritzl00}, putting them in a third Oosterhoff class.  These
features may all be produced by populations of differing He enrichment
\citep{caloidantona07,dantonacaloi08}.

While the HB of NGC 6441 has been studied several times before
\citep{yoon08,busso,caloidantona07,raimondo02,rich97}, two subjects have not
been addressed: 1) although the signature of differential reddening in the
cluster has been seen, researchers have not attempted to {\it remove} its
effect on the CMD, and 2) the spatial distribution of the
stars throughout the cluster has not been examined. If a second generation of
stars forms from gas enriched by a previous generation, and the gas has had a
chance to cool and collect in the gravitational potential of the cluster, the
second generation of stars is expected to have a more concentrated
distribution. This effect seems to have been observed in several clusters, as
in the two MS populations of $\omega$ Cen \citep{bellini09}, 
the giant branch of NGC 3201 \citep{carretta10},
and potentially many others \citep{carretta09}. In a massive and old cluster,
however, dynamical effects like mass segregation may affect the stellar
distribution.  We therefore set out to examine NGC 6441 more closely.

\section{Data}

NGC 6441 is a massive bulge cluster with $M_V=-9.63$ \citep{mclvdm05}.  We
rereduced {\it HST WFPC2} archival data taken as part of proposal IDs 5667,
6095, 8251, and 8718, and {\it ACS HRC} data taken in proposal ID 9835.  These
datasets, which contain optical and some near-UV observations, were selected in
order to obtain nearly complete samples of HB stars over a large fraction of
the cluster's mass.  The data cover from the center of the cluster to a radius
of $\sim$120$\arcsec$, representing $\sim$3 half-light radii ($r_h=34\farcs2$;
\citealt{mclvdm05}), or about 15 core radii ($r_c=7\farcs77$).

\begin{figure}[h]
\begin{center}
\includegraphics[scale=0.6,angle=90,trim=0in 0in 0in 0in]{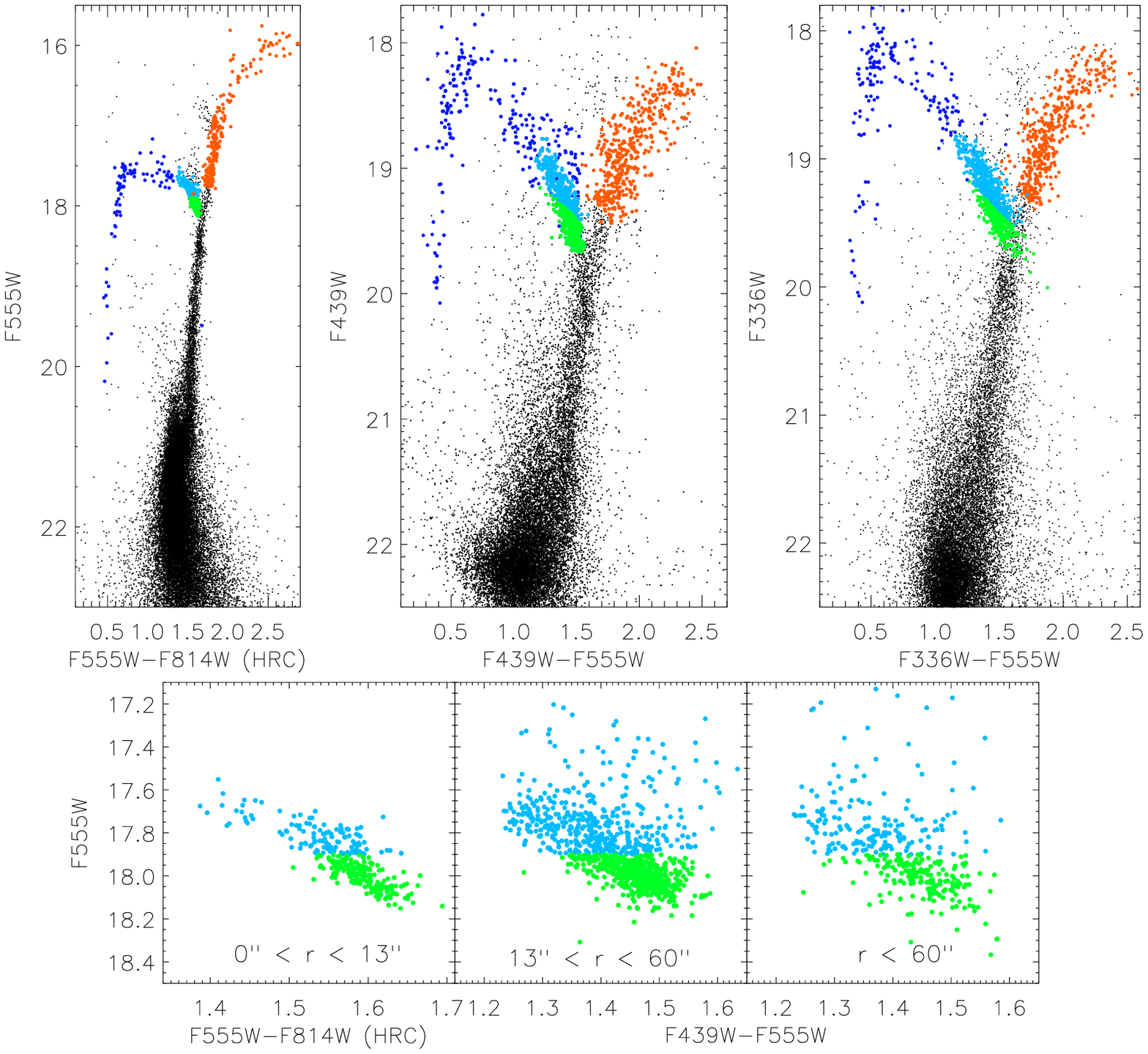}
\caption{The CM diagrams along the top show the populations used in the analysis.  These are the faint red clump (fRC; green), the bright red clump (bRC; light blue), the blue horizontal branch (BHB; dark blue) and red giant branch (RGB; orange).  The leftmost CMD shows data from the ACS HRC while the other two show WFPC2 data.  The three CMDs along the bottom are zoomed in on the red clump at three different radial bins.  The innermost bin uses HRC data and the others WFPC2.}
\label{fig:cmd}
\end{center}
\end{figure}

The WFPC2 images were analyzed using the HSTPhot package \citep{dolphin00},
and stars were cross-identified among the multiple pointings. Median internal
magnitude error estimates were 0.02-0.03 mag at the level of the horizontal
branch for the images (from proposal 8251) covering the largest portion of the
field discussed here. The ACS HRC images were processed using the ACS module
of DOLPHOT \citep{dolphin2002}. Populations were primarily identified via the
color-magnitude diagrams (CMDs) shown in Fig. \ref{fig:cmd} --- primarily 
ACS F555W versus F555W-F814W for the cluster core, and WFPC2 F439W versus
F439W-F555W elsewhere.  A different procedure was needed for fainter blue HB
stars (as discussed below). The HB star group contains 2125,
while the combined red
giant/asymptotic giant (RGB/AGB) group 912
stars.  The HB can be subdivided
into the red clump (RC; 1754
stars) and the ``blue'' HB (BHB; 371 stars).
The identification of the BHB sample deserves some additional explanation. The
sample described here includes RR Lyrae variables (identified from
\citealt{pritzl03} and \citealt{corwin}) and some red HB stars bluer than the
clump, but does not include extreme HB stars ($m_{555} > 20$) as those stars
were found to have significant incompleteness. The BHB stars were validated
using near-UV data from proposal 8718 (PI Piotto) and 5667 (PI Margon; F336W
filter only) when available because crowding of star images is reduced in the
UV where cooler giants and main sequence stars emit less.  UV data was
only available for two mostly overlapping WFPC2 pointings, and so did not
cover the entire extent of the optical data. Importantly though, it does cover
the center of the cluster where blending with brighter cluster stars is the
most likely to cause BHB stars to be lost in optical photometry. In the F255W
filter for example, the blue HB stars are among the brightest cluster sources,
and crowding effects are negligible. Outside of the fields covered by the UV
images, the F439W filter plays a similar role by enhancing the brightess of
BHB stars relative to the numerous faint RGB and MS stars. Our artificial star
tests indicate that incompleteness among our BHB sample is less
than 3\% in the central regions.

\citet{nataf11} and \citet{milone} have shown that structure within the RC of
47 Tuc can also reflect population differences within a cluster. Because the
RC population in NGC 6441 is large, it provides an excellent probe of the
radial structure of the cluster even when subdivided. In light of this, we
examined several different ways of bisecting the RC population: by magnitude,
by line parallel to the slope of the HB, and by line perpendicular to the
slope of the HB.  In this way we can look for signs of population differences
somewhat independent of theoretical expectations. Because the results of
subdividing the clump could be affected by the cluster's differential
reddening (see Fig. 7 of \citealt{raimondo02} or Fig. 4 of \citealt{busso}),
we corrected for this by determining the median color shift of lower RGB stars
($18.3 < m_{555} < 19.2$) from a mean line in the CMD. The correction was
determined for each HB star from RGB stars within $9\arcsec$ or $15\arcsec$
depending on the number of RGB stars available. 
In the center of the cluster, we relied on the high resolution photometry from
the {\it ACS HRC} (proposal 9835, PI Drukier) to reduce incompleteness and
also to minimize the effects of blending in the photometry. If uncorrected,
this allows some clump stars to be measured as too bright and allows some red
giants to masquerade as clump stars when blended with fainter stars.  Our
artificial star tests indicate that our incompleteness for clump stars is less
than 1\% across the entire {\it HRC} field, and the median magnitude offset
between input and output stars was approximately 0.002 mag.

Sky coordinates were determined using the METRIC task in the STSDAS
IRAF package
Before examining the radial distributions of stars in the cluster, 
we recalculated the cluster centroid using a method similar to that
used by \citet{montegriffo95}, and found to be $\alpha = 17^{\rm
  h}50^{\rm m}17\fs9$ and $\delta = -37:03:05.93$, or about $6\arcsec$
from that of \citet{harris96}.

\section{Results}

\begin{figure}[h]
\begin{center}
\includegraphics[scale=0.5,angle=90]{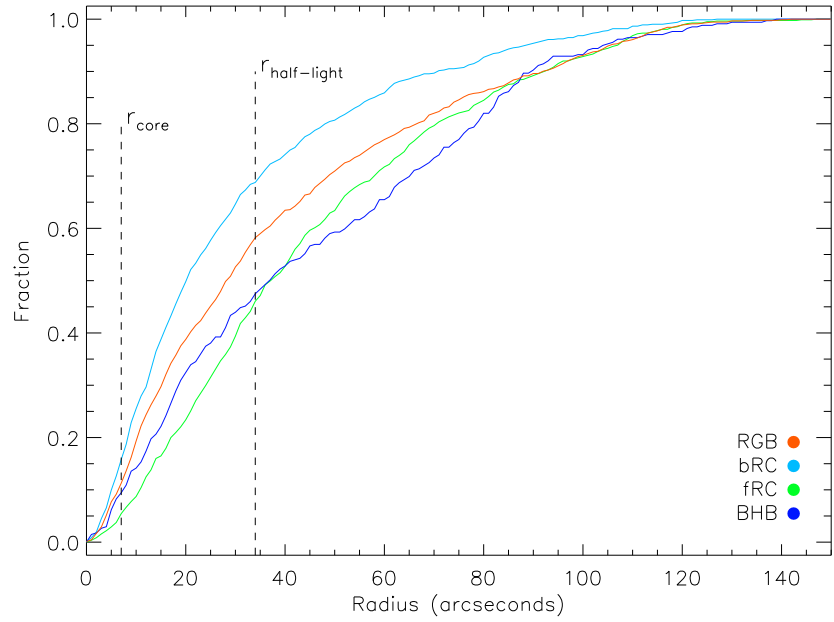}
\caption{Cumulative radial distributions of the analyzed populations.}
\label{fig:crds}
\end{center}
\end{figure}

We used Kolmogorov-Smirnov (KS) test statistics to calculate the probabilities
that the cumulative radial distributions (CRDs) of different subpopulations
were drawn from the same underlying distributions (Fig. \ref{fig:crds}). 
In our analysis, the RGB/AGB group was used as a stand-in for the cluster as a
whole as this portion of the CMD does not show multiple populations as 
clearly.

When examining subdivisions of the red clump, we found that cuts
parallel to the HB did not produce radial distributions of the bright and
faint components that were significantly different.  Cuts perpendicular to the
HB and cuts by F555W magnitude both resulted in groups that had significant
differences in their radial distributions, and the results of KS tests were
essentially equivalent. (This is perhaps expected given the steep slope of the
HB at the location of the RC.) For the lack of a strong reason for selection,
we concentrate on cuts made at a F555W magnitude of 
17.9 for the rest of this
study, and designate the subpopulations the bright red clump (bRC) and faint
red clump (fRC).
As seen in Fig. \ref{fig:crds}, the bRC is quite
significantly more centrally concentrated than the fRC, with the probability
that the two are drawn from the same underlying distribution being $8\e{-17}$.
The bRC is also more centrally concentrated than the RGB/AGB. 

A closer examination of the clump reveals clues to why this is. After the
clump has been corrected for differential reddening, the color extent of the
clump is similar in the three radial bins we selected, indicating that the
issue has been largely removed. In the innermost bin, there are few stars at
the faint red end of the clump ($m_{555} \gtsimeq 17.95$). There is a strong
concentration of stars at that end of the clump in the middle radius bin, but
in the outermost bin the stars are distributed nearly uniformly across the
clump. This behavior can be seen in the CMD array in Fig. 4 of \citet{busso}
as well --- there is a tendency for the clump to appear stubbier and more
concentrated in color outside the centralmost part of the cluster.

The BHB shows yet another pattern in its radial distribution. Overall, the CRD
(Fig. \ref{fig:crds}) indicates that the BHB is less centrally concentrated
than both the bRC and the RGB/AGB.  Comparison of the BHB and RGB/AGB
distributions produces a KS probability (that they derive from the same
underlying radial distribution) of $4\e{-4}$. But it is also interesting that
the slope of the distribution between $\sim$25$\arcsec$ and $\sim$60$\arcsec$,
is lower than those for other populations. The shallower slope perhaps
suggests that the radial distribution of BHB stars is bimodal, though more
weakly than is often seen in blue straggler studies \citep[e.g.]{ferraro}.

\begin{figure}[h]
\begin{center}
\includegraphics[scale=0.5,angle=90]{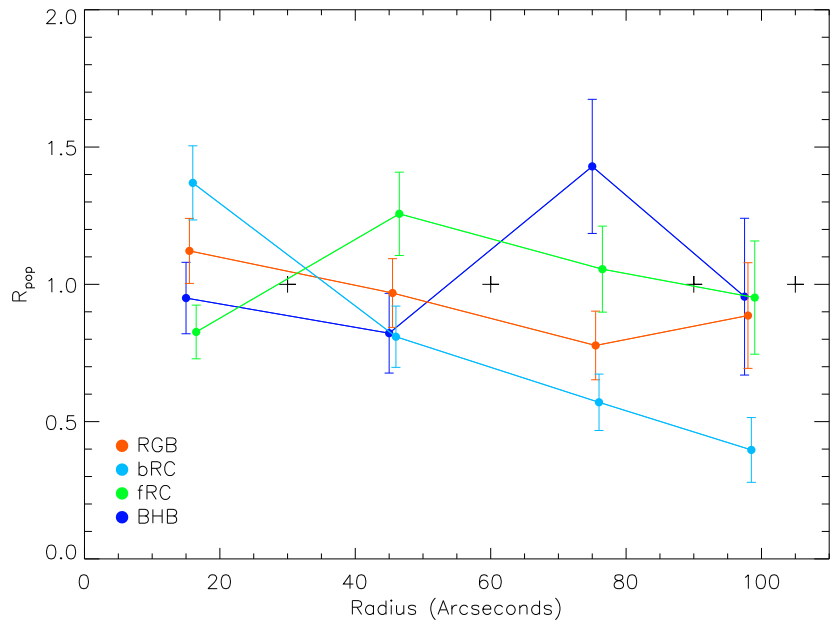}
\caption{Radial distribution of the double-normalized ratios for the selected
  populations. Black crosses mark the boundaries of the chosen annuli.
 The points are offset horizontally from one another for clarity.}
\label{fig:dnrs}
\end{center}
\end{figure}

To clarify the situation further, we have compared the stellar populations
with the cluster light distribution using doubly-normalized ratios $R_{pop} =
(N_{pop} / N_{pop}^{total}) / (L_{sampled} / L_{sampled}^{total})$ from blue
straggler studies \citep{ferraro93}. We calculated the sampled luminosity
$L_{sampled}$ using a King model with structural parameters taken from
\citet{mclvdm05} that fits the surface brightness distribution of the
cluster. The ratio is calculated in annular radial bins centered on the
cluster, and is designed to equal 1 if the stellar population matched the
light distribution (see Fig. \ref{fig:dnrs}).  In the $R_{pop}$ ratios, the
BHB stars show an enhancement in numbers in the third radial bin
($60-90\arcsec$) that is significant at the $2\sigma$ level.  The bRC stars
show the most significant deviations from following the light, with the
population ratio steadily decreasing moving outward from the center.

Before discussing the physical meaning of these distributions, it is worth
commenting on possible complicating factors. Although others have concluded
that differential reddening is not responsible for the slope of the the clump
\citep{busso,raimondo02,swei}, we applied a correction as described in \S 2 to
eliminate the majority of its effect. The referee wondered whether blending
could produce the strong central concentration seen in the bRC
population. This idea is contradicted by the fact that strong central
concentration among the brighter clump stars is still seen when the clump is
split along a line perpendicular to the HB (more like a color cut) while the
concentration disappears when the cut is made parallel to the HB (more like a
magnitude cut). It could also be imagined that incompleteness might be an
issue in the population of BHB stars (particularly among the faint blue tail
in the optical CMD), but there are several ways to see this is a minor
consideration. First, in the core where incompleteness is the biggest issue,
our star selections made use of near-UV photometry --- this reveals the
hottest HB stars even when very close to redder HB stars and giants, and
crowding issues are negligible in the near UV.  Second, the blue tail of the
BHB is a small fraction (about 13\%) of the stars in the BHB sample. So while
we may have lost some ($\sim$3) of the hottest HB stars (faintest in the
optical) to incompleteness, this would not affect conclusions about the
radial distributions.

\section{Discussion}

The impression from the red clump stars in NGC 6441 is that the bluest
and brightest of the red clump stars are considerably more centrally
concentrated than the stars at the faint red end of the clump, which
are found in a small area of the CMD. However, the even hotter BHB
stars are less centrally concentrated than the bRC stars and show a
radial distribution of different character. These indications of {\it
  three} HB populations with differing radial distributions in NGC
6441 are difficult to fit into a consistent picture.

Cluster self-enrichment is the current leading explanation for
multiple populations in globular clusters, involving a second
generation of stars formed from gas polluted in helium and some
heavier elements by ejecta from a group of stars such as
intermediate-mass asymptotic giant branch stars \citep{ventura08} or
rapidly-rotating massive stars \citep{decressin07} in the first
generation (FG).  There are several factors that affect the radial
distribution of the second generation (SG), although one of the
primary predictions is that the formation of the second generation
only begins once gas has cooled and concentrated in the potential well
of the cluster. SG stars are thus expected to be more centrally
concentrated than the FG, although later dynamical evolution of the
cluster could modify this.

\citet{caloidantona07} produced synthetic models of NGC 6441 HB stars
using this picture with the fRC representing FG stars (helium
abundance $Y=0.25$, and having a compact distribution in the CMD),
while the bRC contained SG stars reaching helium abundances of at
least $Y = 0.35$.
BHB stars in the models involved a combination of stars with even
higher abundances (up to $Y = 0.40$) and stars that had evolved from
the bright end of the clump (large helium abundances result in large
blueward excursions during HB evolution). They estimated that the
enriched HB population contained 60\% of the stars, although they
based their study on a smaller sample of stars and did not account for
differential reddening, which smears out the peak at the faint end.
This last factor also influences whether or not a gap 
in $Y$ between FG and SG stars is inferred (see their Fig. 7)

On the basis of the CMD, there is good evidence of two
populations. However, the radial distributions paint a more complex
picture. Specifically, the BHB stars do not not continue the trend
that hotter HB stars (presumably even more helium enriched) are also
centrally concentrated.

Though the second generation may be born with a more centrally concentrated
distribution, simple physics tells us two-body interactions must occur,
causing the SG stars to diffuse outward over time.  This effect would be
enhanced by the fact that helium-enriched stars are expected to be less
massive than stars of normal composition because they can evolve off the main
sequence more quickly \citep{dantonacaloi08}.  Two-body interactions in a
cluster lead the stars toward equipartition of kinetic energy, ``heating'' the
SG stars and giving them a more extended distribution \citep{carretta09}.
This process is expected to act within a Hubble time \citep{dercole08},
affecting the innermost part of the distribution most quickly. Over time, the
cluster core should begin to be depleted of the {\it progenitors} of BHB
stars, while stars on larger orbits would be less affected.

As an exercise, we can calculate the radius at which the dynamical relaxation
timescale equals the lifetime of the cluster.  We again use cluster structural
parameters from \citet{mclvdm05} and the following equation from \citet{bt08},
which describes a stellar system that is a singular isothermal sphere:
\begin{equation}
t_{relax} = \frac{5\, \mbox{Gyr}}{\ln{\Lambda}} \frac{1 M_\odot}{m} \frac{\sigma}{10\, \mbox{km\, s}^{-1}} \left( \frac{r}{1\, \mbox{pc}} \right)^{2}
\label{}
\end{equation}
and find that the relaxation radius is $\sim$60$\arcsec$. This falls within
the part of the BHB radial distribution that has shallower slope than the rest
of the stellar groups. Qualitatively, the BHB distribution shows the
signatures of dynamical modifications.

Other mechanisms such as gas expulsion \citep{decressin10} can affect the
dynamical evolution of the cluster, so that more detailed models will be
needed to verify whether the radial distributions are consistent with expected
masses for helium-enriched populations. Examination of other dynamically
interesting populations, such as blue stragglers, would also clarify the state
of the cluster structure. In the more massive blue straggler star populations
of many clusters, there is a bimodal radial distribution, where the ``zone of
avoidance'' corresponds well to the area expected to be cleared out dynamical
friction.  However, in NGC 6388, a cluster similar to NGC 6441 in many
respects, blue stragglers still populate that area, making the cluster appear
dynamically young \citep{dalessandro08}. 

\acknowledgements We thank the anonymous referee for very helpful comments on
the paper. This research was supported by AST grant 05-07785 from the National
Science Foundation to E.L.S. and M. Bolte.

\end{document}